\documentclass[journal]{IEEEtran}
\usepackage{amsfonts}
\usepackage{xcolor,soul,framed} 
\usepackage{threeparttable}  
\colorlet{shadecolor}{gray!15}
\usepackage[pdftex]{graphicx}
\graphicspath{{../pdf/}{../jpeg/}}
\DeclareGraphicsExtensions{.pdf,.jpeg,.png}
\usepackage{hyperref}
\usepackage[cmex10]{amsmath}
\usepackage{array}
\usepackage{mdwmath}
\usepackage{mdwtab}
\usepackage{eqparbox}
\usepackage{url}
\usepackage{booktabs}
\usepackage{multirow}
\usepackage{siunitx} 
\usepackage{threeparttable}  
\usepackage{textcomp} 
\begin{document}
\bstctlcite{IEEEexample:BSTcontrol}
    \title{Traffic-aware Hierarchical Integrated Thermal and Energy Management for Connected HEVs}
  \author{Jie Han,~\IEEEmembership{ Member,~IEEE,}
          Arash Khalatbarisoltani,~\IEEEmembership{ Member,~IEEE,}
          Hai L. Vu,~\IEEEmembership{ Senior Member,~IEEE,}\\
      Xiaosong Hu,~\IEEEmembership{Fellow,~IEEE,}
      Jun Yang,~\IEEEmembership{Fellow,~IEEE}

  \thanks{This work was funded in part by the 
  Technical Innovation and Application Development Special Program of Chongqing Major Project (CSTB2024TIAD-STX0031), the National Natural Science Foundation of China (No. U23A20327 and 72361137006), and the Basic Research Funds for Central Universities (No. 2023CDJQCZX-001). (\textit{Corresponding author: Jun Yang})}
  \thanks{J. Han and J. Yang are with the Department of Aeronautical and Automotive Engineering, Loughborough University, LE11 3TU Loughborough, U.K. (e-mail: j.han@lboro.ac.uk, j.yang3@lboro.ac.uk).}
  \thanks{A. Khalatbarisoltani and X. Hu is with the Department of Mechanical and Vehicle Engineering, Chongqing University, Chongqing 400044, China (e-mail: arash.khalatbarisoltani@cqu.edu.cn, xiaosonghu@ieee.org).}
  \thanks{H. Vu is with the Department of Civil Engineering, Monash University, Melbourne, VIC 3800, Australia (e-mail: hai.vu@monash.edu).} 
  }  

\markboth{IEEE Transactions on Intelligent Vehicles, 2026
}{Roberg \MakeLowercase{\textit{et al.}}: High-Efficiency Diode and Transistor Rectifiers}

\maketitle
\begin{abstract}
The energy and thermal management systems of hybrid electric vehicles (HEVs) are inherently interdependent. With the ongoing deployment of intelligent transportation systems (ITSs) and increasing vehicle connectivity, the integration of traffic information has become crucial for improving both energy efficiency and thermal comfort in modern vehicles. To enhance fuel economy, this paper proposes a novel traffic-aware hierarchical integrated thermal and energy management (TA-ITEM) strategy for connected HEVs. In the upper layer, global reference trajectories for battery state of charge (SOC) and cabin temperature are planned using traffic flow speed information obtained from ITSs. In the lower layer, a real-time model predictive control (MPC)-based ITEM controller is developed, which incorporates a novel Transformer-based speed predictor with driving condition recognition (TF-DCR) to enable anticipatory tracking of the reference trajectories. Numerical simulations are conducted under various driving cycles and ambient temperature conditions. The results demonstrate that the proposed TA-ITEM approach outperforms conventional rule-based and MPC-SP approaches, with average fuel consumption reductions of 56.36\% and 5.84\%, respectively, while maintaining superior thermal regulation and cabin comfort. These findings confirm the effectiveness and strong generalization capability of TA-ITEM and underscore the advantages of incorporating traffic information.
\end{abstract}

\begin{IEEEkeywords}
Integrated thermal and energy management, traffic information, speed prediction, connected hybrid electric vehicles
\end{IEEEkeywords}

\IEEEpeerreviewmaketitle

\section{Introduction}

\IEEEPARstart{W}{ith} the growing worldwide issues of energy sustainability and climate change, it is essential to accelerate the promotion of low-carbon and ecologically friendly transportation electrification \cite{han2023energy}. Hybrid electric vehicles (HEVs) have emerged as a promising solution for the transition from internal combustion engine vehicles (ICEVs) to pure electric vehicles (EVs) \cite{liang2023efficient}. Given that HEVs constitute a highly coupled thermal and mechanical system, it is necessary to realize the efficient coordination between energy and thermal management subsystems to fully tap into the energy-saving potential \cite{wei2019integrated}. Nevertheless, most existing approaches rely on rule-based strategies \cite{gong2019integrated} or decoupled optimization of thermal and energy domains \cite{he2023review}, which results in suboptimal fuel economy and limits control adaptability and robustness in dynamic real-world driving conditions.

Recent integrated thermal and energy management (ITEM) research has made substantial advances in investigating the impact of component thermal dynamics on vehicular energy consumption. Shams-Zahraei et al. \cite{shams2012integrated} proposed an ITEM strategy based on dynamic programming (DP) for HEVs to optimize both battery energy and engine temperature, which investigated the effects of engine thermal dynamics on vehicular fuel economy and emissions. To balance the real-time implementation and optimum control, Hu et al. \cite{hu2021multihorizon} developed a multihorizon model predictive control (MH-MPC) approach that integrates a short receding horizon using accurate vehicle speed previews and a long shrinking horizon using approximate vehicle speed previews for global cost-to-go estimation. Similarly, Wu et al. \cite{wu2024integrated} introduced a hierarchical MH-MPC approach that optimizes the battery thermal regulation and power flow sequentially, resulting in significant reductions in battery degradation and energy consumption. Despite the excellent energy-saving performance of these ITEM strategies, they mainly focus on onboard optimization and overlook external factors such as traffic conditions, which limits their ability to anticipate and adapt to dynamic driving conditions \cite{yang2023modeling}.

The emergence of intelligent transportation systems (ITSs) and vehicle connectivity has enabled HEVs to access traffic information (e.g. future speed profiles, road gradients, and traffic signal phases). Integrating traffic information has opened a new optimization dimension for predictive ITEM control. Zhao et al. \cite{zhao2021two} designed a two-layer ITEM strategy, where the upper-level controller first optimizes battery temperature trajectories based on a priori speed preview, while the lower-level controller performs energy consumption optimization by tracking the planned thermal trajectories. Amini et al. \cite{amini2019cabin} presented a hierarchical MPC approach incorporating both short-term vehicle speed prediction and long-term traffic condition prediction. This approach also utilized a novel intelligent online constraint handling (IOCH) method to optimize cabin and battery cooling strategies. However, the power split ratio wasn't optimized in real time, which was only determined by a rule-based strategy. Dong et al. \cite{dong2024predictive} introduced a predictive battery thermal and energy management (p-BTEM) strategy, combining DP-based global trajectory planning in the cloud controller with the MPC-based real-time energy optimization in the onboard controller. Furthermore, advanced artificial intelligence techniques such as deep reinforcement learning (DRL) have also been employed for smart ITEM control. Zhang et al. \cite{zhang2023integrated} proposed a TD3-based ITEM approach that integrates multi-source traffic and terrain data to enhance energy-saving performance via coordination engine warm-up and heating strategies. Khalatbarisoltani et al. \cite{khalatbarisoltani2025privacy} developed a federated reinforcement learning framework that enables collaborative ITEM across multiple vehicles while preserving data privacy.

However, a common limitation among existing methods lies in the insufficient integration of traffic foresight into the lower-level control loop \cite{lokur2023distributed,amini2020hierarchical}. Specifically, predicted speed profiles or traffic information are typically leveraged in upper-level planners to generate reference trajectories, while the real-time controllers often rely on static setpoints or predefined rules, lacking adaptability to dynamic driving conditions \cite{gong2019integrated,hu2022multirange}. These limitations constrain the control ability to respond effectively to real-world traffic fluctuations and diminish the potential benefits of predictive control. Recent advances in deep learning offer promising tools to address this gap. High-fidelity data-driven speed predictors, particularly those based on advanced architectures such as Recurrent Neural Networks (RNNs) \cite{lu2021combined}, Long Short-Term Memory (LSTM) networks \cite{pulvirenti2023energy}, and Transformer \cite{zhu2025vspnet}, have shown remarkable accuracy in forecasting future vehicle states. When integrated with MPC, these predictors significantly enhance the anticipatory capability and robustness of real-time decision-making, enabling proactive coordination between thermal and energy domains under dynamic and uncertain environments \cite{quan2021real}.

While the integration of traffic foresight and predictive speed profiles has shown promising potential in enhancing energy and thermal management, most existing approaches overlook the critical influence of driving condition categories on vehicle speed dynamics \cite{zhou2022comprehensive}. Each driving condition exhibits distinct statistical patterns and temporal dependencies. For instance, urban driving is characterized by frequent stop-and-go behavior and high variability, highway driving typically features smoother and more stable speed trajectories, while suburban conditions lie between these two extremes with moderate fluctuations \cite{lin2021velocity}. Ignoring these differences can lead to suboptimal prediction accuracy and diminished control robustness, especially in naturalistic driving environments where transitions between driving modes are common \cite{li2020driver,li2025driver}.

To address this challenge, this study proposes a novel traffic-aware hierarchical integrated thermal and energy management (TA-ITEM) strategy that improves real-time control performance in fuel economy and cabin comfort. Specifically, in the upper layer, the global reference trajectories of battery state of charge (SOC) and cabin temperature are optimized using traffic flow speed. Then, an MPC-based ITEM framework is employed by integrating a Transformer-based speed prediction model with driving condition recognition (TF-DCR) to explicitly identify current driving conditions and adapt the prediction model accordingly. The main contributions of this study are summarized as follows:
\begin{itemize}
\item A control-oriented thermal model is developed and validated against high-fidelity simulations, achieving an optimal balance between computational efficiency and modeling accuracy.
\item A novel TF-DCR speed predictor is proposed, capable of better capturing the influence of driving style on vehicle speed variation and yielding significantly improved prediction accuracy over traditional approaches.
\item A novel traffic-aware hierarchical ITEM strategy is designed and validated, demonstrating consistent enhancements in fuel efficiency and cabin comfort across diverse naturalistic driving scenarios.
\end{itemize}

The remainder of this paper is organized as follows: Section~\ref{se:Modeling} presents the powertrain and thermal management system models. Section~\ref{se:Traffic_data} details the traffic dataset and traffic flow speed extraction methodology. Section~\ref{se:Traffic-aware ITEM} introduces the TA-ITEM strategy and the TF-DCR speed predictor. Simulation results and comparative analyses are provided in Section~\ref{se:Results and discussion}, followed by conclusions and future work directions in Section~\ref{se:Conclusion}.

\section{Modeling of Powertrain and Thermal Management Systems} \label{se:Modeling}
\begin{figure}
  \begin{center} \includegraphics[width=0.48\textwidth]{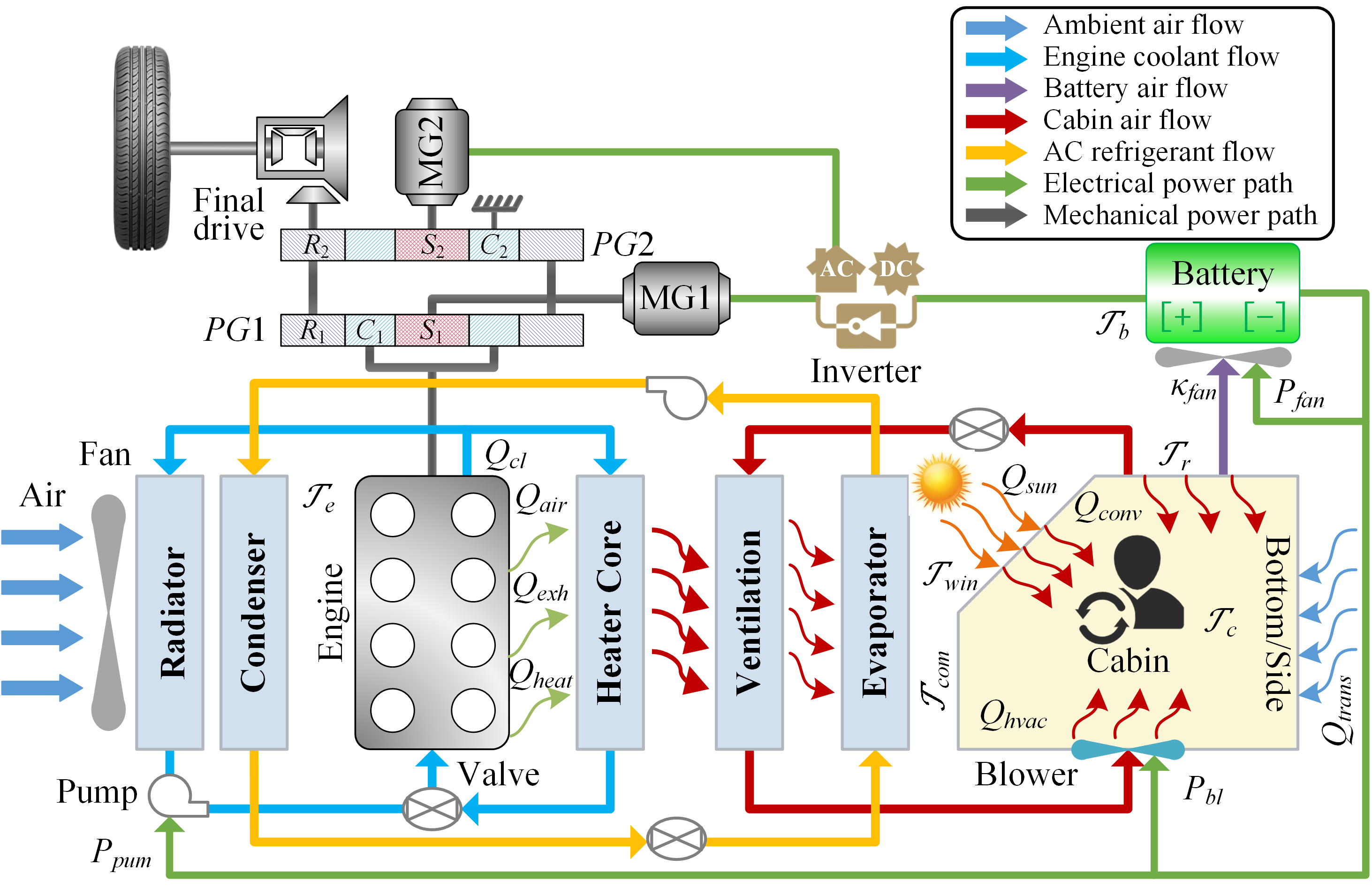}\\
  \caption{Schematic of the powertrain and thermal management system of Prius HEV.}\label{fig:Powertrain_TMS}
  \end{center}
\end{figure}

The powertrain dynamics and thermal management systems of the Prius HEV are presented in this section, as depicted in Fig.~\ref{fig:Powertrain_TMS}. The parameters of the vehicle and the powertrain \cite{Autonomie2024} are listed in Table~\ref{tab:vehicle_parameters}.

\sisetup{
  per-mode=symbol,
  separate-uncertainty=true,
  detect-all
}

\begin{table}[htbp]
  \centering
  \caption{Vehicle and Powertrain Parameters}
  \label{tab:vehicle_parameters}
  \begin{tabular}{lll}
    \toprule
    \text{Category} & \text{Parameter} & \text{Value} \\
    \midrule

    \multirow{8}{*}{Vehicle} 
    & Vehicle mass $m_v$ & \SI{1350}{\kilogram} \\
    & Gravity constant $g$ & \SI{9.8}{\meter\per\square\second} \\
    & Air density $\rho_{air}$ & \SI{1.184}{\kilogram\per\cubic\meter} \\
    & Rolling resistance $c_r$ & 0.007 \\
    & Air drag coefficient $c_d$ & 0.3 \\
    & Front area $A_f$ & \SI{1.746}{\square\meter} \\
    & Tire radius $r_w$ & \SI{0.28}{\meter} \\
    & Wheel efficiency $\eta_w$ & 0.95\\   
    \midrule
    
    \multirow{2}{*}{Power-split}
    & PG1 ratio $r_1$ & 78/30 \\
    & PG2 ratio $r_2$ & 2.5 \\
    \midrule

    \multirow{2}{*}{Engine}
    & Max. power $P_{e,max}$ & \SI{110} {\kilo\watt}\\
    & Max. torque $T_{e,max}$ & \SI{220} {\newton \meter}\\
    \midrule

    \multirow{2}{*}{MG1}
    & Max. power $P_{mg1,max}$ & \SI{40} {\kilo\watt}\\
    & Max. torque $T_{mg1,max}$ & \SI{170} {\newton\meter}\\
    \midrule

    \multirow{2}{*}{MG2}
    & Max. power $P_{mg2,max}$ & \SI{60} {\kilo\watt}\\
    & Max. torque $T_{mg2,max}$ & \SI{207} {\newton\meter}\\
    \midrule
    
    \multirow{3}{*}{Battery cell}
    & Nominal voltage $u_{nom}$ & \SI{1.2}{\volt} \\
    & Nominal capacity $Q_{nom}$ & \SI{6.5}{\ampere\hour} \\
    & Cell number $n_b$ & 168  \\
    \midrule

    \multirow{2}{*}{Final drive}
    & Gear ratio $r_d$ & 3.26 \\
    & Efficiency $\eta_d$ & 0.97 \\
    \bottomrule
  \end{tabular}
\end{table}

\subsection {Longitudinal Dynamics}
The vehicle longitudinal dynamics \cite{onori2016hybrid} are expressed in \eqref{eq:vehicle_dynamics},
\begin{equation}\label{eq:vehicle_dynamics}
    F_v = m_v \cdot g (c_r \cdot\cos{\theta} + \sin{\theta}) + 0.5 \rho_{air} \cdot c_d\cdot A_f\cdot v^2 + m_v\cdot a 
\end{equation}
where $F_v$ is the demanded force for a given driving cycle, $m_v$ is the vehicle mass, $g$ is the gravity constant, $c_r$ is the rolling resistance coefficient, $\theta$ is the road slope, $\rho_{air}$ is the aerodynamic drag coefficient, $A_f$ is the vehicle front area,  $v$ is the vehicle speed, and $a$ is the acceleration. Then the output torque $T_{ps}$ and speed $\omega_{ps}$ of the power-split device are calculated by \cite{bonab2020fuel}
\begin{equation}
    T_{ps} = \frac{F_v \cdot r_w}{r_d \cdot (\eta_d \cdot \eta_w)^{\operatorname{sign}(F_v)}}
\end{equation}
\begin{equation}
    \omega_{ps} = \frac{v \cdot r_d}{r_w}
\end{equation}
where $r_w$ is the wheel radius, $r_d$ is the final drive gear ratio, and $\eta_d$ and $\eta_w$ denote the transmission efficiencies of the final drive and the wheel, respectively.

\subsection {Power-split Device}
In the Prius powertrain, there are two planetary gears (PGs) that connect the engine, motor/generator 1 (MG1), and motor/generator 2 (MG2). As shown in Fig. \ref{fig:Powertrain_TMS}, the engine
and MG1 are connected to the planet carrier and the sun gear of PG1, respectively. In PG2, the planet carrier is held stationary and the MG2 is connected to the sun gear. In addition, the ring gears of both PG1 and PG2 are interconnected and coupled to the final drive. Quasi-static equations for the power-split device are \cite{han2022joint}
\begin{equation}
\left\{
\begin{aligned}
    &T_{r1} = \frac{r_1 \cdot T_e}{r_1 + 1} = -r_1 \cdot T_{mg1} \\
    &\omega_{mg1} = (r_1 + 1) \cdot \omega_e - r_1 \cdot \omega_{ps} \\
    &\omega_{mg2} = r_2 \cdot \omega_{ps} \\
    &r_2 \cdot T_{mg2} = T_{ps} - T_{r1}
\end{aligned}
\right.
\end{equation}
where $r_1$ and $r_2$ are the gear ratios of PG1 and PG2, $T$ and $\omega$ represent the torque and speed, and the subscript $r_1$, $e$, $mg1$, and $mg2$ refer to the ring gear of PG1, engine, MG1, and MG2, respectively. 

\subsection {Engine Model}
The engine fuel consumption is modeled as a lookup table with engine torque and speed as inputs. The instantaneous fuel consumption rate $\dot{m}_f$ is
\begin{equation}
    \dot{m}_f = \delta_f \cdot f_f(\omega_e,T_e)
\end{equation}
\begin{equation}
    P_e = \omega_e \cdot T_e
\end{equation}
where $P_e$ is the engine mechanical power, $\delta_f$ is the factor of additional fuel consumption reflected by the engine temperature.

In this paper, the engine coolant temperature is treated as the engine temperature and is expressed as
\begin{equation}
    \dot{\mathcal{T}}_e = \frac{1}{m_e \cdot C_e} (Q_f - P_e - Q_{exh} - Q_{air} - Q_{cl} - Q_{heat})
\end{equation}
where $m_e$ represents the equivalent thermal mass of the engine, $C_e$ is its equivalent specific heat capacity, $Q_f$ denotes the total heat released from fuel combustion, $Q_{exh}$, $Q_{air}$, and $Q_{cl}$ represent the heat dissipated through exhaust gas, engine compartment air, and coolant, respectively, and $Q_{heat}$ is the heat transferred to the cabin. Then the heat dissipation can be calculated by \eqref{eq:heat_dissipation}.

\begin{equation} \label{eq:heat_dissipation}
\left\{
\begin{aligned}   
&Q_f = \delta_h \cdot \dot{m}_f \cdot Q_{lhv}\\
&Q_{exh} = \delta_{exh} \cdot (Q_f - P_e)\\
&Q_{air} = h_e \cdot A_e \cdot (\mathcal{T}_e - \mathcal{T}_{com})\\
&Q_{cl} = f_{cl}(\mathcal{T}_e, \mathcal{T}_{amb}, W_{e,cl}, W_{e,air})\\
\end{aligned}
\right.
\end{equation}
where $\delta_h$ is the heat ratio by engine temperature, $\delta_{exh}$ is the heat added ratio by air temperature, and $Q_{lhv}$ is the low heating value of the fuel. $h_e$ is the heat transfer coefficient of the engine, $A_e$ is the equivalent heat transfer area, $\mathcal{T}_{com}$ refers to the engine compartment temperature, and $f_{cl}$ represents the heat exchange model of the radiator, which depends on engine temperature, ambient temperature $\mathcal{T}_{amb}$, coolant flow rate $W_{e,cl}$, and intake air flow rate $W_{e,air}$.

\subsection {Motor/Generator Model}
In this subsection, the efficiency of MG~$\eta_{mg}$ is characterized as a function of torque and speed.
\begin{equation}
\left\{
\begin{aligned}
&P_{mg} = \omega_{mg} \cdot T_{mg}\\
&P_{mg,dc} = \frac{P_{mg}}{\eta_{mg}(\omega_{mg}, T_{mg})^{\operatorname{sign}(P_{mg})}}
\end{aligned}
\right.
\end{equation}
where $P_{mg}$ and $P_{mg,dc}$ are the mechanical power and electric power of MG, respectively.

\subsection {Battery Model}
Ref. \cite{han2023health} demonstrated that a battery model coupling the internal resistance model with a lumped-mass thermal model provides a satisfactory balance between computational efficiency and simulation accuracy under room temperature conditions. Therefore, the internal resistance model is adopted to capture the electrical dynamics.
\begin{equation}
 u = u_{oc} - i \cdot R_0 \label{eq:bat_voltage} \\   
\end{equation}
\begin{equation}
 i = \frac{u_{oc} - \sqrt{u_{oc}^2 - 4 R_0 \cdot P_{cell}}}{2 R_0} \label{eq:bat_current} \\   
\end{equation}
\begin{equation}
P_{cell} = \frac{P_{mg1,dc} + P_{mg2,dc} + P_{pump} + P_{fan} + P_{bl}}{n_{{b}}} \label{eq:Pcell} \\  
\end{equation}
\begin{equation}
\dot{soc} = \frac{-i}{3600 \cdot Q_{nom}} \label{eq:soc_dynamics}    
\end{equation}
where $u_{oc}$ is the open-circuit voltage, $u$ is the terminal voltage, $R_0$ refers to the internal resistance. $i$ represents the cell current, and $P_{cell}$ represents the output power of the cell.  $n_b$ is the number of cells in the battery pack, $Q_{nom}$ is the nominal capacity of the cell, and $\dot{soc}$ is the state of charge (SOC) rate. Additionally, $P_{pump}$, $P_{fan}$, and $P_{bl}$ are the power consumption of the engine coolant pump, the battery cooling fan, and the HVAC ventilation blower, respectively.

The generated thermal $Q_b$ and thermal dynamics $\dot{\mathcal{T}}_b$ of the battery can be expressed as 
\begin{equation}
\left\{
\begin{aligned}
     &Q_b = i^2 \cdot R_0 \cdot n_b\\
    &\dot{\mathcal{T}}_b = \frac{1}{m_b \cdot C_b}(Q_b - Q_{cool})\\   
\end{aligned}
\right.
\end{equation}
where $m_b$ is the equivalent thermal mass of the battery and $C_b$ is its equivalent specific heat capacity. $Q_{cool}$ represents the heat dissipated through the battery cooling system and can be calculated by
\begin{equation}
 \begin{gathered}
Q_{cool} = f_{cool}(\mathcal{T}_b,\mathcal{T}_c,W_{b,air}(\kappa_{fan},v))\\
\end{gathered}   
\end{equation}
where $f_{cool}$ represents the function of the battery cooling system, which depends on battery temperature, cabin temperature $\mathcal{T}_c$, cooling air flow rate $W_{b,air}$, and $\kappa_{fan}$ is the fan duty cycle of the battery cooling system.

\subsection {Cabin Thermal Model}
In this paper, cabin thermal comfort is also considered as one of the optimization objectives. The cabin temperature rise rate can be expressed as
\begin{equation}
    \dot{\mathcal{T}_c} = \frac{1}{m_c \cdot C_c + C_{add}}(Q_{hvac} + Q_{sun} + Q_{trans} + Q_{conv})
\end{equation}
where $m_c$ is the equivalent thermal mass of the cabin, $C_c$ denotes the specific heat capacity of the cabin, and $C_{add}$ represents the additional heat capacity contributed by other materials in the cabin. The heat flows $Q_{hvac}$, $Q_{sun}$, $Q_{trans}$, and $Q_{conv}$ represent the heat input/output from the HVAC system, solar radiation, heat transmission through the cabin structure, and convective heat exchange, respectively, which can be calculated by:
\begin{equation}
\left\{
 \begin{aligned}
&Q_{sun} = \eta \cdot I \cdot \beta \cdot A_{w}\\
&Q_{trans} = (h_{c,b} \cdot A_{c,b} + h_{c,s} \cdot A_{c,s}) \cdot (\mathcal{T}_{amb} - \mathcal{T}_c)\\
&Q_{conv} = -Q_{c2w} - Q_{c2r}\\
&Q_{c2w} = h_w \cdot A_w \cdot (\mathcal{T}_c - \mathcal{T}_w)\\
&Q_{c2r} = h_r \cdot A_r \cdot (\mathcal{T}_c - \mathcal{T}_r)\\
\end{aligned}  
\right.
\end{equation}
where \( \eta \) is the radiation transmission coefficient of the glass, \( I \) is the solar radiation intensity, and \( \beta \) is the perpendicular component of solar radiation on the window. \( A_w \), \( A_r \), \( A_{c,b} \), and \( A_{c,s} \) denote the areas of the window, roof, cabin bottom, and cabin side surfaces, respectively. \( h_{c,b} \), \( h_{c,s} \), \( h_w \), and \( h_r \) are the heat transfer coefficients of the corresponding surfaces. \( Q_{c2w} \) and \( Q_{c2r} \) represent the convective heat flows from the windows and the roof to the cabin air. \( \mathcal{T}_w \) and \( \mathcal{T}_r \) denote the temperatures of the window and roof surfaces.

\section{Traffic Information Description}\label{se:Traffic_data}

\subsection {Traffic Dataset}\label{se:Traffic Information}
\begin{figure}
  \begin{center}
  \includegraphics[width=0.45\textwidth]{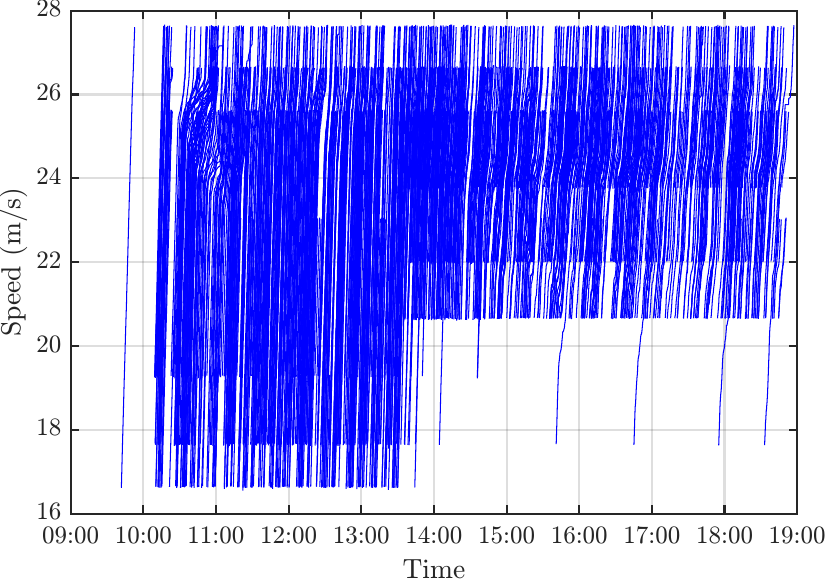}
  \caption{Individual vehicle trajectories for northbound trips collected during the Mobile Century field experiment conducted on the I-880 highway \cite{herrera2010evaluation}.}
  \label{fig:Vehicle_trajectory}
  \end{center}
\end{figure}

The deployment of large-scale traffic monitoring systems, such as the performance measurement system (PeMS) in California \cite{chen2001freeway}, enables vehicle users to access historical and real-time traffic data to improve vehicular energy consumption. However, traditional traffic monitoring methods (e.g. inductive loop detectors and surveillance cameras) are costly and difficult to maintain. In contrast, GPS-enabled smartphones have emerged as a promising alternative due to their widespread adoption and high-precision positioning capabilities. The Mobile Century experiment \cite{herrera2010evaluation} collected trajectory data from 100 individual vehicles equipped with GPS-enabled Nokia N95 smartphones along California's I-880 highway on February 8, 2008 (10:00–18:00 PST). The collected 1388 vehicle trajectories are shown in Fig.~\ref{fig:Vehicle_trajectory}. 

\begin{figure}
  \begin{center}  \includegraphics[width=0.45\textwidth]{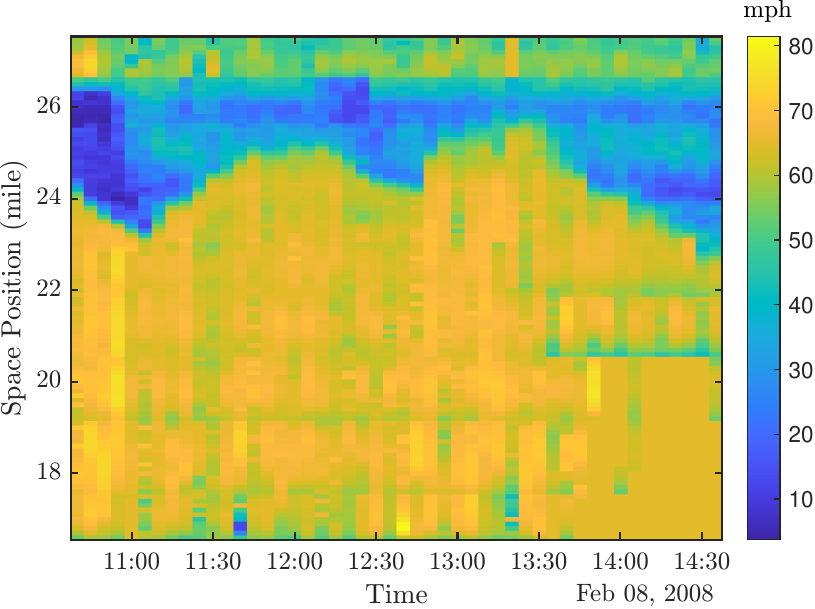}
  \caption{The average spatio-temporal traffic flow speed.}
  \label{fig:Traffic_speed}
  \end{center}
\end{figure}

To reconstruct the spatiotemporal traffic flow speed, the vehicle trajectory data were aggregated over a predefined time-space grid. The observation period (10:40–14:40) was divided into 5-minute intervals (the same as the updating frequency of PeMS \cite{caltrans_pems}), and the highway segment (mile 16.7–27.6) was divided into 0.1-mile intervals \cite{6930758}. For each vehicle, the instantaneous speed records were mapped to their corresponding time-space cells, and the mean speed per cell was calculated. The average traffic speed \( \bar{v}_{s,t} \) in spatial cell \( s \) and temporal interval \( t \) was then obtained by aggregating speed measurements from all vehicles as follows:
\begin{equation}
\bar{v}_{s,t} = \frac{1}{N_{s,t}} \sum_{i=1}^{N_{s,t}} v_{i,s,t}
\end{equation}
where \( N_{s,t} \) denotes the number of vehicle speed records within the \((s,t)\) cell, and \( v_{i,s,t} \) is the instantaneous speed of the \( i \)-th sample. For cells without data, the average speed is set to the road speed limit to maintain matrix consistency and prevent artificial congestion signals caused by missing data. The spatio-temporal average traffic flow speed is illustrated in Fig.~\ref{fig:Traffic_speed}.

\subsection{Traffic Flow Speed Extraction}\label{se:speed extraction}
To extract the traffic flow speed along a specific vehicle trajectory, a dynamic interpolation method is applied over a discretized spatio-temporal speed field \( \bar{v}(s,t) \). The vehicle's initial time and position \((t_0, s_0)\) are first determined from the raw data. Then, at each integration step, the traffic speed \( \bar{v}_i = \bar{v}(s_i, t_i) \) is obtained from the average traffic speed matrix. Based on this local speed, the vehicle is assumed to traverse a small spatial cell \( \Delta s \), and the corresponding time increment \( \Delta t \) is computed as:
\begin{equation}
 \Delta t_i = \frac{\Delta s_i}{\bar{v}_i}   
\end{equation}
The vehicle position and time are iteratively updated as:
\begin{equation}
s_{i+1} = s_i + \Delta s_i, \quad t_{i+1} = t_i + \Delta t_i  
\end{equation}
until the terminal time or position is reached. The reconstructed velocity profile is then interpolated to match the sampling resolution of the original trajectory for subsequent control. This method ensures that the extracted traffic flow speed \( v_{tra}(t) \) accurately reflects the spatial-temporal distribution of traffic conditions experienced by each individual vehicle. 

\begin{figure}
  \begin{center}  \includegraphics[width=0.45\textwidth]{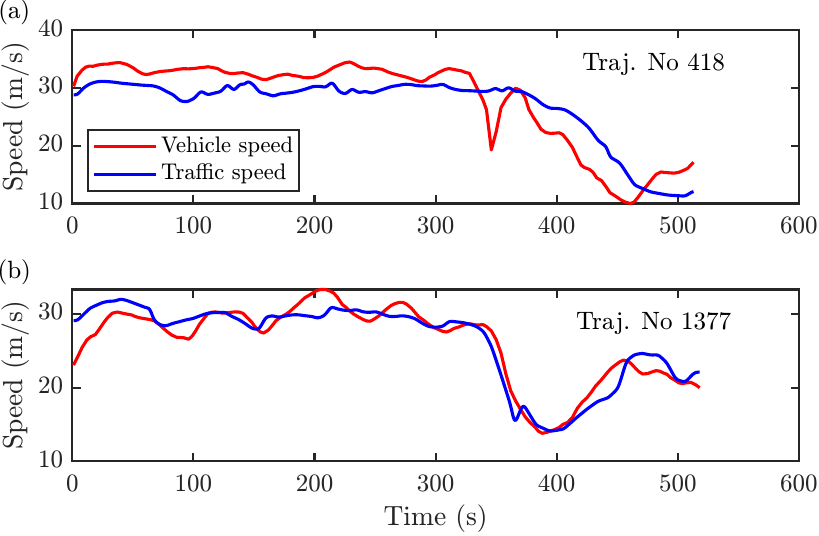}
  \caption{Comparison between extracted traffic flow speed and individual vehicle speed for two trajectories. (a) No.418 and (b) No. 1377.} \label{fig:Traffic_flow_speed}
  \end{center}
\end{figure}

Two vehicle trajectories were randomly selected, and the corresponding traffic flow speeds were extracted using the aforementioned method. As illustrated in Fig. ~\ref{fig:Traffic_flow_speed}, the extracted traffic flow speeds can effectively capture the overall trend of vehicle speed variations. Therefore, in Section \ref{se:Global Trajectory Planning}, the traffic flow speed is utilized to plan the trajectories of battery SOC and cabin temperature, thereby overcoming the limitations of conventional MPC-based control strategies, particularly their inability to fully exploit the entire SOC range.

\section{Traffic-aware Hierarchical Integrated Thermal and Energy Management} \label{se:Traffic-aware ITEM}
\begin{figure*}[ht]
  \centering  \includegraphics[width=0.6\textwidth]{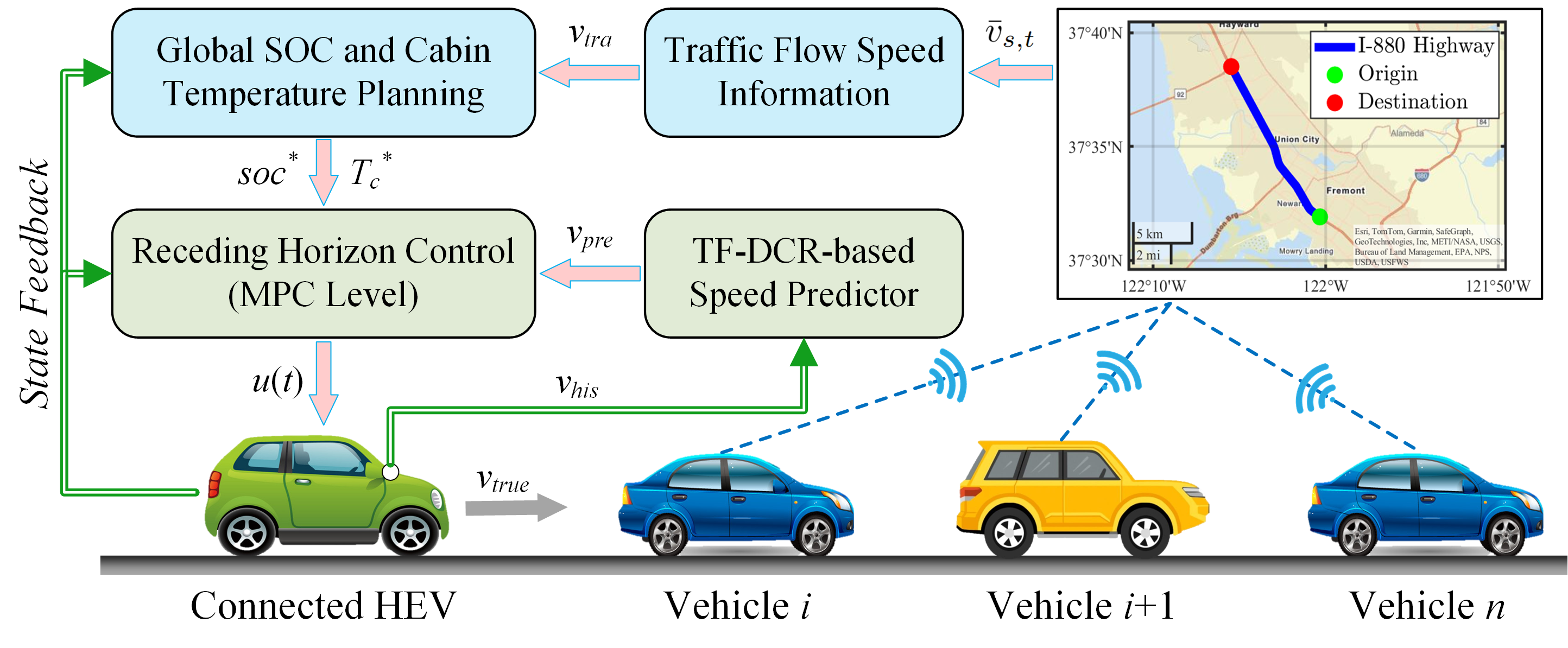}
  \caption{Control framework for traffic-aware hierarchical integrated thermal and energy management.}
  \label{fig:Control framework}
\end{figure*}

In this section, a novel traffic-aware hierarchical ITEM strategy is proposed as shown in Fig. ~\ref{fig:Control framework}. In the upper layer, the optimal reference trajectories of battery SOC $soc^*$ and cabin temperature $\mathcal{T}_c^*$ are planned by utilizing the traffic flow speed extracted as described in Section~\ref{se:speed extraction}. In the lower layer, a real-time ITEM is implemented based on MPC, which integrates TF-DCR to track the reference trajectories.

\subsection {Global Reference Trajectory Planning Formulation}\label{se:Global Trajectory Planning}
In this subsection, the global ITEM strategy is developed based on traffic flow speed \(v_{tra}\), aiming to minimize total fuel consumption while ensuring cabin comfort and SOC level. The control actions include engine speed and torque, the battery cooling fan duty cycle, and the HVAC heat flow.
\begin{equation}
    \mathbf{u} = [\omega_e,T_e,\kappa_{fan},Q_{hvac}]
\end{equation}
The state variables consist of battery SOC, and the temperatures of battery, cabin air, roof surface, windows, engine coolant, and engine compartment.
\begin{equation}
    \mathbf{x} = [soc,\mathcal{T}_b,\mathcal{T}_c,\mathcal{T}_r,\mathcal{T}_w,\mathcal{T}_e,\mathcal{T}_{com}]
\end{equation}
The corresponding optimal control problem (OCP) minimizes the sum of the instantaneous fuel consumption over the whole horizon and a penalty term weighted by 
$\alpha$, which accounts for deviations from the target cabin temperature.
\begin{equation}
\begin{aligned}
    \min_{\mathbf{u}(t)} &\quad \sum_{i=1}^{N} \dot{m}_f(t) + \alpha \cdot \delta_{\mathcal{T}_c}(t)\\
    s.t 
    &\quad 0 \leq \kappa_{fan} (t) \leq 1\\
    &\quad Q_{hvac,min} \leq Q_{hvac} (t) \leq Q_{hvac,max}\\
    &\quad \omega_{j,min}(t) \leq \omega_{j} (t) \leq \omega_{j,max}(t)\\
    &\quad T_{j,min}(t) \leq T_{j} (t) \leq T_{j,max}(t)\\
    &\quad \mathbf{x}_0(\text{2:7})=\mathcal{T}_{amb}\\
    &\quad soc(0) = soc(N)\\
    &\quad  soc_{min} \leq soc(t) \leq soc_{max}\\
    &\quad \mathcal{T}_{b,min}  \leq \mathcal{T}_b(t) \leq \mathcal{T}_{b,max}\\
    &\quad \mathcal{T}_{c,tar}-\delta_{\mathcal{T}_c}(t)  \leq \mathcal{T}_c(t) \leq \mathcal{T}_{c,tar}+\delta_{\mathcal{T}_c}(t)\\
\end{aligned}
\end{equation}
where $N$ is the total control horizon length. The variable $\delta_{\mathcal{T}_c}$ represents the allowable deviation of the cabin air temperature $\mathcal{T}_c(t)$ from the desired comfort temperature $\mathcal{T}_{c,tar}$, forming a time-varying comfort constraint $[\mathcal{T}_{c,tar} - \delta_{\mathcal{T}_c}(t),\ \mathcal{T}_{c,tar} + \delta_{\mathcal{T}_c}(t)]$. The initial state vector $\mathbf{x}_0$ includes key thermal variables such as battery, cabin, and engine-related temperatures, which are initialized as the ambient temperature $\mathcal{T}_{amb}$. $soc(0) = soc(N)$ aims to maintain the battery SOC level. The cabin temperature $\mathcal{T}_c(t)$ is thus regulated throughout the control horizon to stay within the predefined comfort range while coordinating with the energy-saving objectives. Considering the complexity and nonlinearity of the OCP, the IPOPT solver in the CasADi toolbox \cite{Andersson2019} is employed to solve the optimization problem efficiently.

\subsection{TF-DCR Speed Predictor}
\begin{figure}
  \begin{center}  \includegraphics[width=0.48\textwidth]{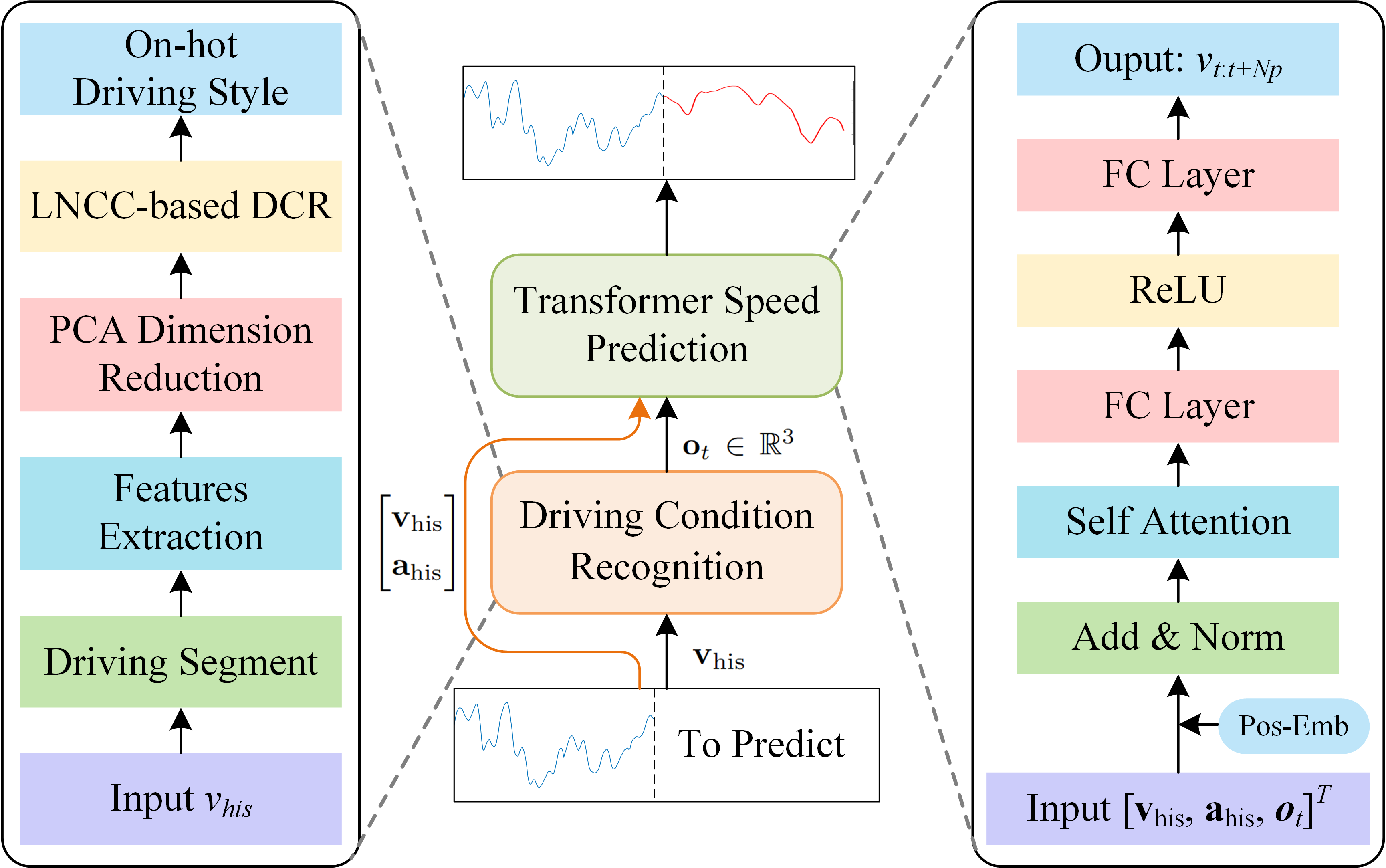}
  \caption{Speed prediction model integrating driving condition recognition and Transformer}
  \label{fig:Speed predictor}
  \end{center}
\end{figure}

In real-world driving scenarios, vehicle speed profiles exhibit substantial variability due to diverse traffic conditions and driver behaviors, such as stop-and-go traffic, steady cruising, or aggressive driving in congested environments. These behavioral patterns correspond to distinct driving condition styles that significantly influence the dynamics and predictability of vehicle speed. To address this issue, as illustrated in Fig.~\ref{fig:Speed predictor}, a novel speed prediction model, referred to as TF-DCR, is proposed by integrating unsupervised driving condition recognition (DCR) with a Transformer-based architecture. This speed prediction model is further incorporated into the MPC-based ITEM control framework to enhance prediction accuracy and decision-making robustness.

\subsubsection{Unsupervised Driving Condition Recognition}\label{se:DCR}
To effectively identify driving conditions, an unsupervised classification framework based on Gaussian Mixture Models (GMM) is developed. The training dataset \cite{han2023health} comprises eight standard driving cycles along with two naturalistic cycles selected from the Vehicle Energy Dataset (VED) \cite{oh2020vehicle}. The standard driving cycles are sourced from \textit{Autonomie}, a vehicle simulation platform developed by the Vehicle \& Mobility Systems Department at Argonne National Laboratory \cite{Autonomie2024}.

\begin{figure}
  \begin{center}  \includegraphics[width=0.4\textwidth]{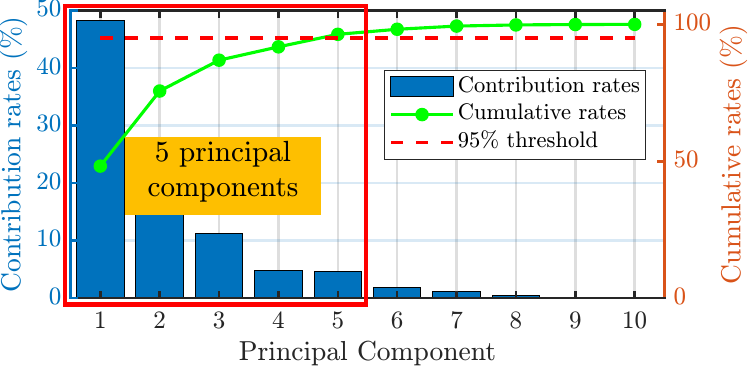}
  \caption{Principal component analysis results.}
  \label{fig:PCA}
  \end{center}
\end{figure}

As a preprocessing step, the training dataset is segmented using a 60-second time window. For each segment, a total of ten statistical features are extracted, including maximum speed, minimum speed, average speed, speed standard deviation, average acceleration, average deceleration, acceleration standard deviation, idle proportion, cruise proportion, and acceleration proportion. Then, Principal Component Analysis (PCA) is applied for feature dimensionality reduction. As illustrated in Fig.~\ref{fig:PCA}, the first five principal components account for over 95\% of the cumulative contribution rates, which are retained for subsequent clustering.

A GMM-based clustering algorithm is subsequently applied to classify driving condition segments into three distinct driving styles: urban, highway, and city. The GMM assumes that the input feature vector \( \mathbf{z} = \phi(\mathbf{v}) \in \mathbb{R}^d \) is generated from a mixture of \( K \) multivariate Gaussian distributions:

\begin{equation}
p(\mathbf{z}) = \sum_{k=1}^{K} \pi_k \, \mathcal{N}(\mathbf{z} \mid \boldsymbol{\mu}_k, \boldsymbol{\Sigma}_k)
\label{eq:gmm}
\end{equation}
where \( \phi(\cdot) \) denotes the feature extraction function, including a PCA process applied to the raw driving data \( \mathbf{v} \). \( d \) is the dimension of the extracted feature vector \( \mathbf{z} \), \( K \) is the total number of Gaussian components, \( \pi_k \) is the non-negative mixing weight of the \( k \)-th component satisfying \( \sum_{k=1}^{K} \pi_k = 1 \); \( \boldsymbol{\mu}_k \) is the mean vector, and \( \boldsymbol{\Sigma}_k \) is the covariance matrix of the \( k \)-th Gaussian component. The term \( \mathcal{N}(\mathbf{z} \mid \boldsymbol{\mu}_k, \boldsymbol{\Sigma}_k) \) represents the multivariate Gaussian distribution, which is defined as:

\begin{equation}
\mathcal{N}(\mathbf{z} \mid \boldsymbol{\mu}, \boldsymbol{\Sigma}) = \frac{1}{(2\pi)^{\frac{d}{2}} |\boldsymbol{\Sigma}|^{\frac{1}{2}}} \exp\left( -\frac{1}{2} (\mathbf{z} - \boldsymbol{\mu})^\top \boldsymbol{\Sigma}^{-1} (\mathbf{z} - \boldsymbol{\mu}) \right)
\label{eq:gaussian}
\end{equation}

To improve numerical stability and avoid overfitting, the regularized GMM is implemented using MATLAB's \texttt{fitgmdist} function.

\begin{figure*}[ht]
  \centering  \includegraphics[width=\textwidth]{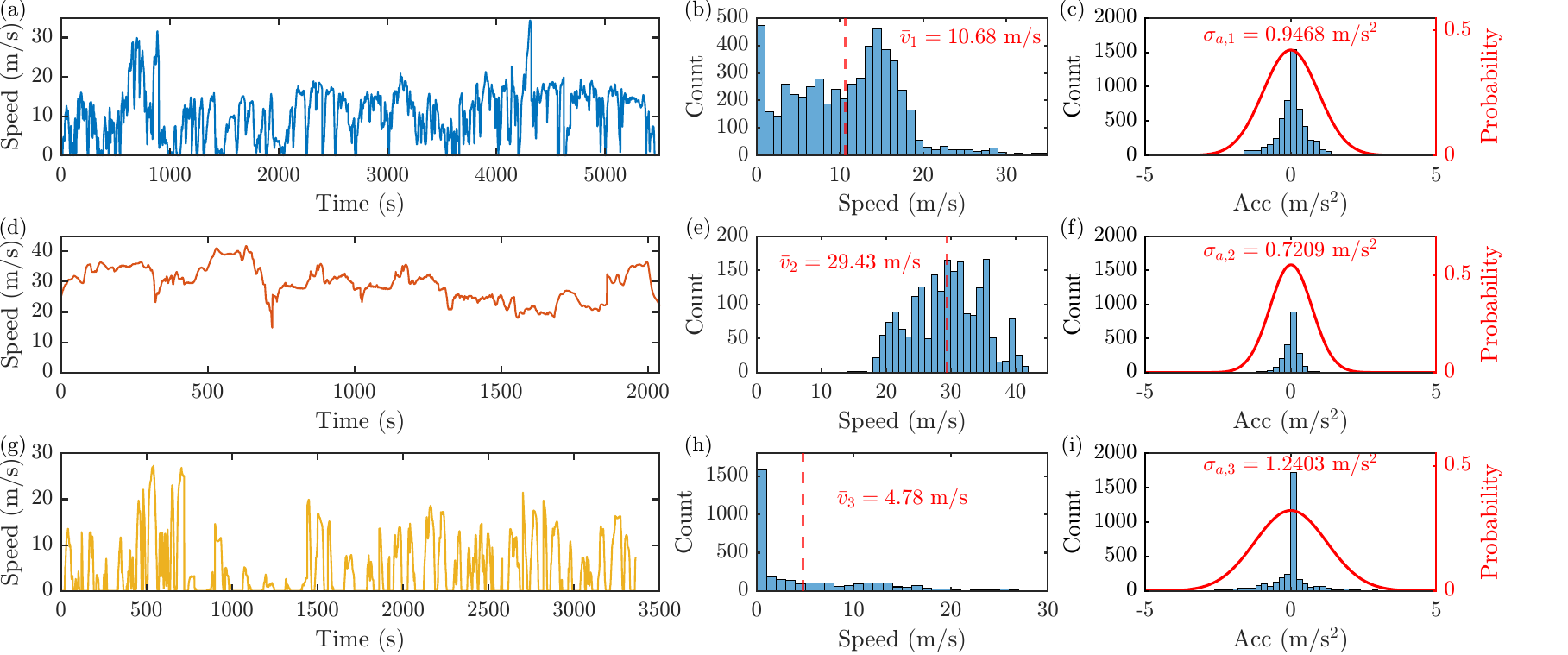}
  \caption{Speed and acceleration time series and distributions under three representative driving conditions: (a)-(c) Urban, (d)-(f) Highway, and (g)-(i) City.}
  \label{fig:Driving style}
\end{figure*}

Fig.~\ref{fig:Driving style} illustrates the clustering results through the distributions of speed and acceleration for each recognized category. The average speeds for these three driving styles are 10.68 m/s, 29.43 m/s, and 4.78 m/s, respectively. Correspondingly, the standard deviations of acceleration are 0.9468 m/s$^2$, 0.7209 m/s$^2$, and 1.2403 m/s$^2$. These distinct statistical characteristics of driving styles validate the effectiveness and rationality of the clustering approach.

To enable real-time recognition, a lightweight nearest-cluster classification (LNCC) approach is employed. Specifically, a test feature vector \( \mathbf{z} \in \mathbb{R}^d \), after normalization and PCA projection, is compared against the \( K \) GMM cluster centers \( \{ \boldsymbol{\mu}_k \} \), and the driving condition style is recognized according to the index of the closest cluster mean:

\begin{equation}
c_t = \arg\min_{k \in \{1, \dots, K\}} \left\| \mathbf{z} - \boldsymbol{\mu}_k \right\|_2
\label{eq:min_dist}
\end{equation}
where \( c_t\) denotes the recognized driving condition style, and the Euclidean distance \( \| \cdot \|_2 \) is used to determine the nearest cluster center. This classification approach reduces computational overhead and is well suited for online inference scenarios.

\begin{figure}[t]
  \centering  \includegraphics[width=0.48\textwidth]{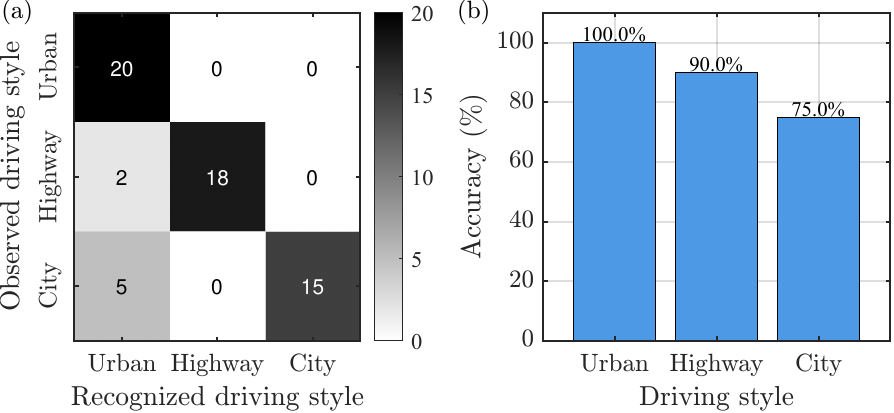}
  \caption{Performance evaluation of the driving condition recognition. (a) Confusion matrix; (b) Recognition accuracy for each driving style.}
  \label{fig:DCR_results}
\end{figure}

For evaluation, 20 driving segments are randomly selected from each driving style. As shown in Fig.~\ref{fig:DCR_results}, the recognition results demonstrate high accuracy, with urban, highway, and city driving condition styles achieving 100\%, 90\%, and 75\% accuracy, respectively. The confusion matrix and sample statistics further verify the effectiveness of the proposed DCR model.

\subsubsection{Transformer-Based Speed Prediction Model}

To enhance the accuracy and robustness of vehicle speed forecasting across diverse traffic environments, a driving condition-aware Transformer framework is proposed. The key idea is to incorporate driving scenario semantic into the prediction model. Specifically, the driving condition for each historical observation window is first identified using a GMM-based clustering approach, and the recognized style label is then embedded and integrated with the sequence input to the Transformer model.

Let \( N_{his} \) and \( N_{pre} \) denote the historical window length and the prediction horizon, respectively. At time \( t \), the historical driving data, including speed and acceleration, are defined as:
\begin{equation}
\begin{aligned}
\mathbf{v}_{\mathrm{his}} &= [v_{t - N_{\mathrm{his}} + 1}, \dots, v_t], \\
\mathbf{a}_{\mathrm{his}} &= [a_{t - N_{\mathrm{his}} + 1}, \dots, a_t].
\end{aligned}
\end{equation}

To embed scenario semantics, the current driving condition style \( c_t \in \{\mathrm{Urban}, \mathrm{City}, \mathrm{Highway}\} \) is identified through the GMM-based clustering and LNCC framework described in Section~\ref{se:DCR}. The label \( c_t \) is encoded as a one-hot vector \( \mathbf{o}_t \in \mathbb{R}^3 \) and broadcast across the temporal dimension.
The normalized input features are then constructed as:
\begin{equation}
\mathbf{X}_t = 
\begin{bmatrix}
\tilde{\mathbf{v}}_{t - N_{his} + 1 : t} \\
\tilde{\mathbf{a}}_{t - N_{his} + 1 : t} \\
\mathbf{o}_t^{1} \\
\mathbf{o}_t^{2} \\
\mathbf{o}_t^{3}
\end{bmatrix}
\in \mathbb{R}^{5 \times N_{his}},
\end{equation}
where \( \tilde{\mathbf{v}} \) and \( \tilde{\mathbf{a}} \) represent normalized speed and acceleration sequences, and \( \mathbf{o}_t^{i} \) denotes the \( i \)-th repeated channel of the one-hot label.

The constructed sequence \( \mathbf{X}_t \) is then processed by a causal Transformer encoder that utilizes multi-head self-attention to model temporal dependencies. The output is projected to the future predicted speed over the next \( N_{pre} \) steps:
\begin{equation}
\hat{\mathbf{v}}_{t+1 : t+N_{pre}} = f_{Trans}(\mathbf{X}_t)
\end{equation}
where \( f_{Trans}(\cdot) \) represents the Transformer network. The model is trained by minimizing the mean squared error (MSE) between the predicted and actual vehicle speeds. The key configuration parameters of the Transformer-based architecture are summarized in Table~\ref{tab:transformer_config}.

\begin{table}
\centering
\caption{Configuration of the Transformer Speed Predictor}
\label{tab:transformer_config}
\renewcommand{\arraystretch}{1.1}
\begin{tabular}{lll}
\toprule
\textbf{Component} & \textbf{Parameter} & \textbf{Value} \\
\midrule
Input features & $[ \tilde{\mathbf{v}}, \tilde{\mathbf{a}},\mathbf{o}_t^{1},\mathbf{o}_t^{2},\mathbf{o}_t^{3}]$ & 5 \\
Historical window size & $N_{his}$ & 60 \\
Prediction horizon & $N_{pre}$ & 5 \\
Position embedding & Maximum sequence length & 256 \\
Self-attention & Number of heads & 4 \\
Self-attention & Key dim per head & 32 \\
Attention mask & Type & Causal \\
Fully connected layer 1 & Output dimension & 128 \\
Activation function & Type & ReLU \\
Fully connected layer 2 & Output dimension & 5 \\
Optimizer & Training algorithm & Adam \\
Initial learning rate & Value & 0.001 \\
Mini-batch size & Training batch & 128 \\
Training epochs & Max iterations & 50 \\
\bottomrule
\end{tabular}
\end{table}

\subsection{MPC-based Integrated Thermal and Energy Management}
To overcome the energy-saving potential limitation caused by hard constraints on battery SOC and cabin temperature, a nonlinear MPC strategy is proposed to track the globally optimal battery SOC and cabin temperature profiles. These reference trajectories are generated in the upper layer by incorporating traffic information. The proposed MPC formulates the control problem as a constrained nonlinear optimization problem over a finite prediction horizon, aiming to achieve optimal trade-offs between energy consumption and thermal comfort while satisfying system dynamics and operational constraints.

The objective function is designed to minimize a weighted sum of tracking errors and fuel consumption.
\begin{equation}
\begin{aligned}
    \min_{\mathbf{u}(\ell \mid t)} \quad J&= \sum_{\ell=t}^{t + N_{pre}-1} 
    \dot{m}_f(\ell \mid t)\\
    &+ \sum_{\ell=t+1}^{t + N_{pre}} w_{soc} \left(soc(\ell \mid t) - soc^*(\ell \mid t)\right)^2\\
    &+ \sum_{\ell=t+1}^{t + N_{pre}} w_{c} \left(\mathcal{T}_c(\ell \mid t) - \mathcal{T}_c^*(\ell \mid t)\right)^2\\
    s.t 
    \quad\quad & 0 \leq \kappa_{fan} (\ell \mid t) \leq 1\\
    \quad\quad & Q_{hvac,min} \leq Q_{hvac} (\ell \mid t) \leq Q_{hvac,max}\\
    \quad\quad & \omega_{j,min}(\ell \mid t) \leq \omega_{j} (\ell \mid t) \leq \omega_{j,max}(\ell \mid t)\\
    \quad\quad & T_{j,min}(\ell \mid t) \leq T_{j} (\ell \mid t) \leq T_{j,max}(\ell \mid t)\\
    \quad\quad & \mathbf{x}_0(\text{2:7})=\mathcal{T}_{amb}\\
    \quad\quad &  soc_{min} \leq soc(\ell \mid t) \leq soc_{max}\\
    \quad\quad & \mathcal{T}_{b,min}  \leq \mathcal{T}_b(\ell \mid t) \leq \mathcal{T}_{b,max}\\
    \quad\quad & \mathcal{T}_{c,min}(\ell \mid t)  \leq \mathcal{T}_c(\ell \mid t) \leq \mathcal{T}_{c,max}(\ell \mid t)\\
\end{aligned}
\end{equation}
where $\ell$ is the prediction step, $t$ is the current time step, $soc(\ell \mid t)$ and $\mathcal{T}_c(\ell \mid t)$ are the predicted battery SOC and cabin temperature at time $\ell$ based on information at time $t$.
$soc^*(\ell \mid t)$ and $\mathcal{T}_c^*(\ell \mid t)$ are the corresponding global reference trajectories from the upper layer. $w_{soc}$ and $w_c$ are the weighting factors balancing SOC tracking and cabin thermal comfort. 

\section{Results and Discussion}\label{se:Results and discussion}

\subsection{Model verification of Thermal Management System}
\begin{figure}
  \centering  \includegraphics[width=0.48\textwidth]{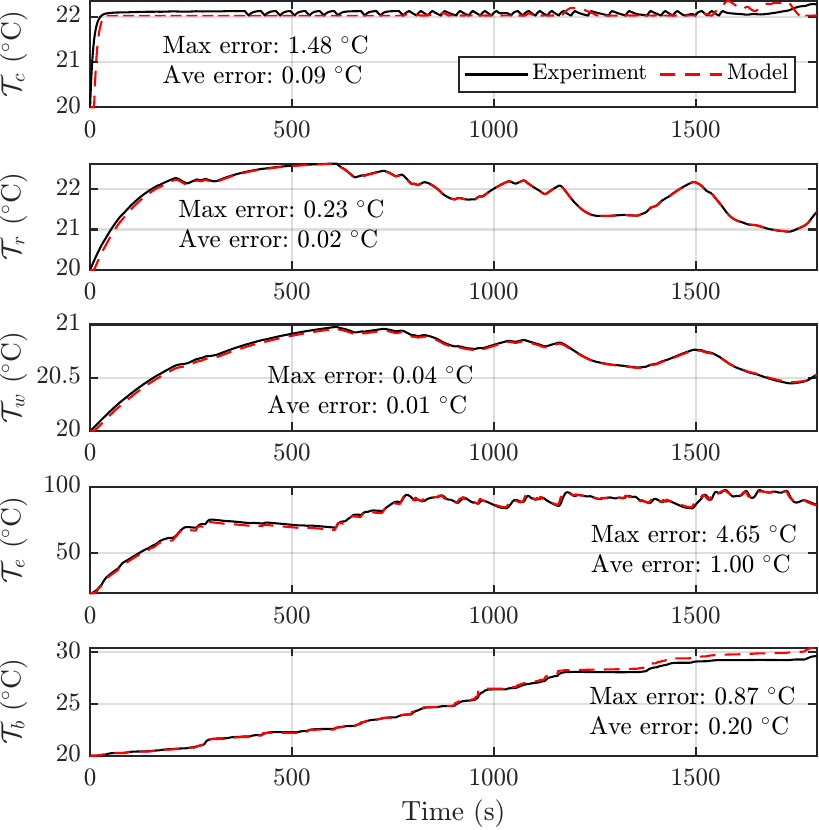}
  \caption{The validation of thermal management model under WLTC cycle.}
  \label{fig:Thermal Model Validation}
\end{figure}

The control-oriented thermal management model is validated against reference data obtained from \textit{Autonomie} under the WLTC driving cycle. As shown in Fig.~\ref{fig:Thermal Model Validation}, the predicted temperatures of the cabin air, roof, window glass, engine, and battery exhibit a strong agreement with the reference data, with an average estimation error within 1 $^\circ$C. This high level of consistency across all thermal subsystems demonstrates the model’s ability to accurately capture both transient and steady-state thermal behaviors, confirming its suitability for real-time MPC applications in ITEM control.

\subsection{Speed Prediction Analysis}
Table~\ref{tab:Speed_prediction performance} presents a comparison of speed prediction performance across various driving cycles using three models: extreme learning machine (ELM), backpropagation neural network (BP), and the proposed TF-DCR. The results demonstrate that TF-DCR consistently outperforms the baseline models, achieving the lowest root mean square error (RMSE) in all scenarios. Notably, TF-DCR achieves up to a 53.17\% improvement over ELM on the No.~1283 vehicle trajectory and shows substantial gains ranging from 9\% to 43\% across most driving cycles. These results highlight the superior generalization ability and robustness of the proposed method under both standard and naturalistic driving conditions, making it highly suitable for real-time speed prediction in modern vehicle applications.

\begin{table}
  \centering
  \caption{Speed Prediction Performance Comparison under Different Driving Cycles}
  \label{tab:Speed_prediction performance}
  \setlength{\tabcolsep}{6pt}  
  \renewcommand{\arraystretch}{1.0}  
  \begin{tabular}{lllc}
    \toprule
    \textbf{Driving Cycle} & \textbf{Predictor} & \textbf{RMSE} & \textbf{Improvement}\\
    \midrule
    \multirow{3}{*}{ARB02}
    & ELM & 2.5464 & -\\
    & BP &  2.1294 & 16.38\%\\
    & TF-DCR & 1.7422 & 31.58\%\\
    \midrule   
    \multirow{3}{*}{UDDS} 
    & ELM & 1.3611 & -\\
    & BP &  1.3058 & 4.06\%\\
    & TF-DCR & 1.1482 & 15.64\%\\
    \midrule
    \multirow{3}{*}{HWFET} 
    & ELM & 0.6415 & -\\ 
    & BP &  0.5906 & 7.93\%\\
    & TF-DCR & 0.5080 & 20.81\%\\
    \midrule   
    \multirow{3}{*}{Traj. No 418} 
    & ELM & 0.7094 & -\\
    & BP &  0.6683 & 5.79\%\\
    & TF-DCR & 0.6418 & 9.53\%\\
    \midrule
    \multirow{3}{*}{Traj. No 724} 
    & ELM & 0.8964 & -\\  
    & BP &  0.6168 & 31.19\%\\
    & TF-DCR & 0.5081 & 43.32\%\\
    \midrule
    \multirow{3}{*}{Traj. No 1283}  
    & ELM & 2.3899 & -\\
    & BP &  1.6607 & 30.51\%\\
    & TF-DCR & 1.1191 & 53.17\%\\
    \bottomrule
  \end{tabular}\\
   \vspace{4pt}
  \raggedright
  \footnotesize \textit{ELM}: Extreme learning machine-based predictor \cite{tang2020naturalistic}; \textit{BP}: Backpropagation neural network-based predictor \cite{han2022predictive}; \textit{TF-DCR}: Transformer-based predictor with driving condition recognition.
\end{table}

\subsection{MPC-based ITEM Strategy}
\begin{figure}
  \begin{center}  \includegraphics[width=0.48\textwidth]{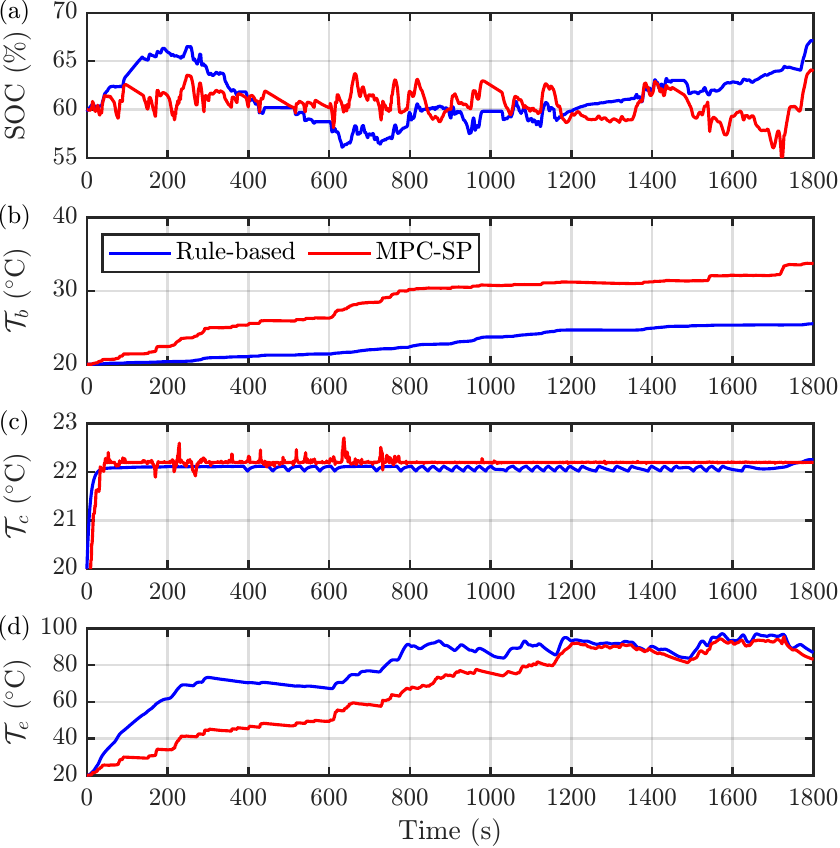}
  \caption{Comparison of rule-based and MPC-SP ITEM strategies under the WLTC cycle.}
  \label{fig:Autonomie_MPC}
  \end{center}
\end{figure}

\begin{figure}
  \begin{center}  \includegraphics[width=0.42\textwidth]{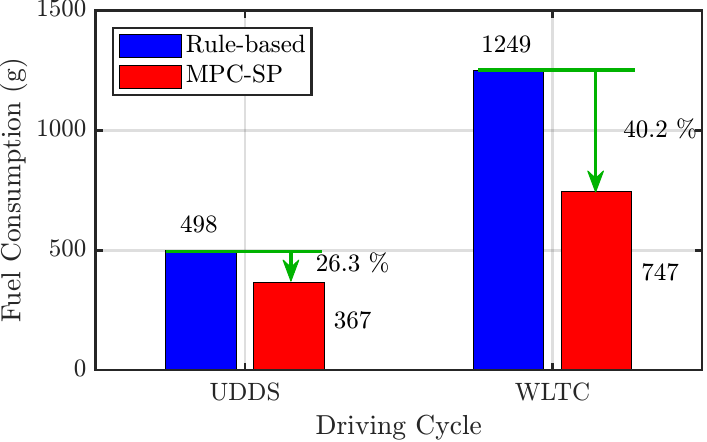}
  \caption{Fuel consumption under UDDS and WLTC cycles.}
  \label{fig:Autonomie_MPC_FC}
  \end{center}
\end{figure}

To implement the real-time ITEM control, the proposed TF-DCR speed predictor is integrated into the MPC-based control strategy, referred to as MPC-SP. Fig.~\ref{fig:Autonomie_MPC} compares the performance of the rule-based approach (implemented in \textit{Autonomie}) and the MPC-SP ITEM strategy under the WLTC driving cycle. It can be seen that the MPC-SP strategy employs more frequent battery utilization, resulting in a higher battery temperature as shown in Fig.~\ref{fig:Autonomie_MPC}(b), approaching the optimal battery temperature threshold \cite{dong2024predictive}. Although the cabin temperature exhibits fluctuations within the first 0–800~s, it rapidly converges to the desired setpoint. Meanwhile, the increased battery usage reduces engine operations, leading to a lower engine temperature.
Furthermore, as illustrated in Fig.~\ref{fig:Autonomie_MPC_FC}, the MPC-SP strategy achieves up to 26.3\% and 40.2\% reductions in fuel consumption under the UDDS and WLTC cycles, respectively, which highlights the superior capability of MPC in jointly optimizing energy efficiency and thermal comfort.

\subsection{Traffic-aware Hierarchical ITEM strategy} 
\begin{figure*}
  \centering  \includegraphics[width=0.9\textwidth]{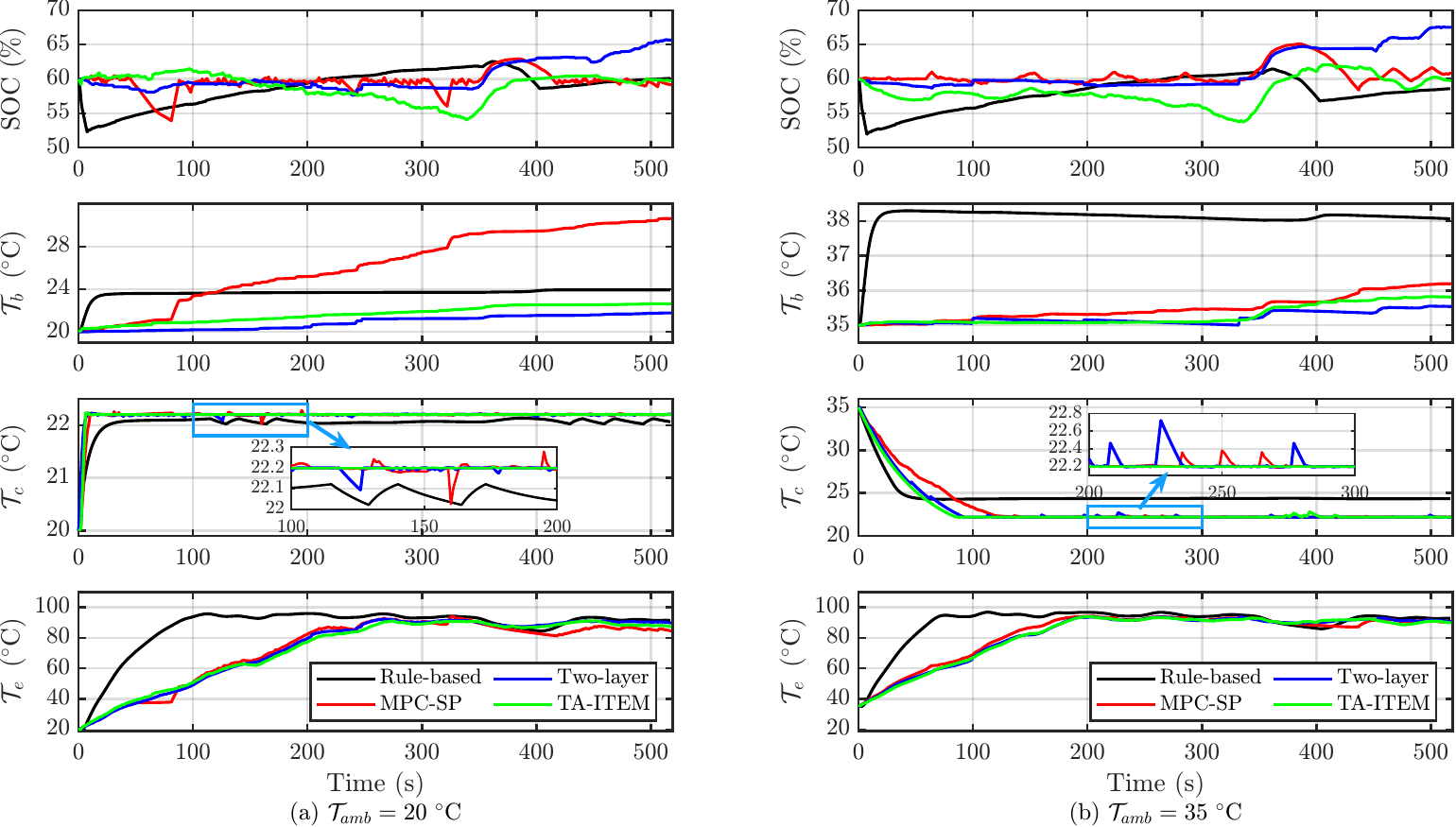}
  \caption{Comparison of different control strategies under No. 1377 vehicle trajectory. (Two-layer strategy is from \cite{zhao2021two}) (a) $\mathcal{T}_{amb}=20~^\circ\mathrm{C}$ and (b) $\mathcal{T}_{amb}=35~^\circ\mathrm{C}$.}
  \label{fig:Control Results}
\end{figure*}

Fig.~\ref{fig:Control Results} compares the dynamic responses of battery SOC, battery temperature, cabin temperature, and engine temperature under different control strategies at ambient temperatures of 20 $^\circ$C and 35 $^\circ$C. Compared to rule-based, MPC-SP, and two-layer control strategies, the proposed TA-ITEM method maintains battery SOC trajectories closer to predefined reference values while expanding the SOC operating window, thereby enhancing the energy buffer for power distribution optimization. Regarding thermal management, TA-ITEM enables faster convergence of cabin temperature to setpoints with smaller steady-state deviations. Furthermore, compared to rule-based and MPC-SP strategies, the TA-ITEM results in reduced battery temperature rise, particularly maintaining closer to the optimal thermal operating point at 35 $^\circ$C, which benefits both efficiency and battery lifespan. Additionally, through more effective battery utilization and load balancing, TA-ITEM reduces engine operating temperatures, thereby indirectly lowering thermal load and fuel consumption.

\begin{figure*}
  \centering  \includegraphics[width=0.8\textwidth]{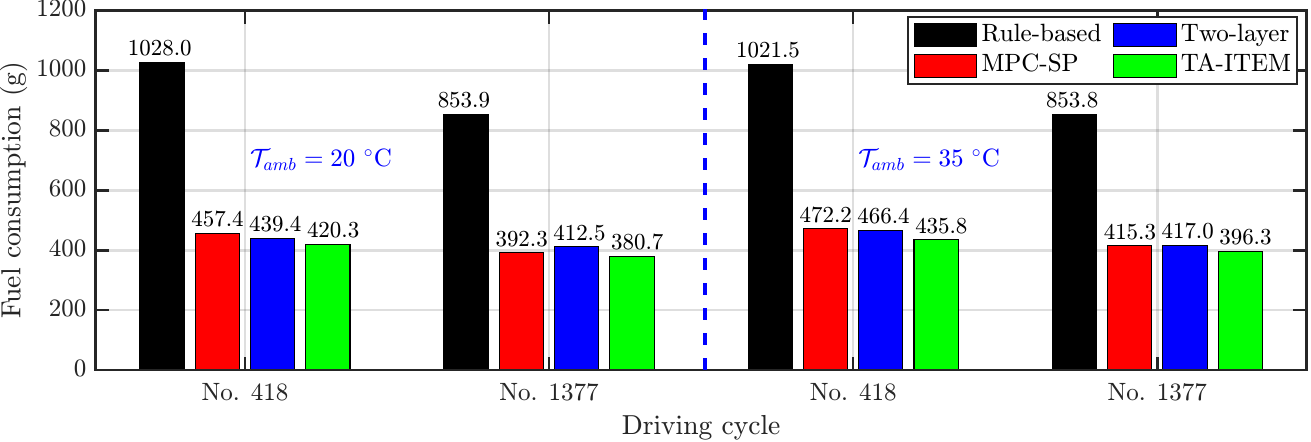}
  \caption{Fuel consumption under different vehicle trajectories and ambient temperatures.}  \label{fig:Control_results_FC}
\end{figure*}

As shown in Fig.~\ref{fig:Control_results_FC}, the proposed TA-ITEM strategy consistently delivers the lowest fuel consumption across all cases. Specifically, under the No.~1377 cycle at 20 $^\circ$C, TA-ITEM achieves up to 55.42\%, 2.96\%, and 7.71\% fuel reductions relative to the rule-based,  MPC-SP, and two-layer strategies, respectively. Moreover, at 35 $^\circ$C, the fuel reductions are 53.58\%, 4.58\%, and 4.96\% respectively. These results clearly demonstrate that TA-ITEM not only improves fuel efficiency but also outperforms the MPC-SP strategy. The superior performance of TA-ITEM can be attributed to its traffic-aware ability. By leveraging traffic information to plan reference SOC and cabin temperature, TA-ITEM allows for a more flexible and optimal scheduling of battery usage and engine operation, thereby improving overall fuel economy. Furthermore, the improvement becomes more pronounced at elevated ambient temperatures, highlighting the robustness of TA-ITEM in mitigating increased cooling demands and maintaining energy-thermal balance.

\begin{figure*}
  \centering  \includegraphics[width=0.8\textwidth]{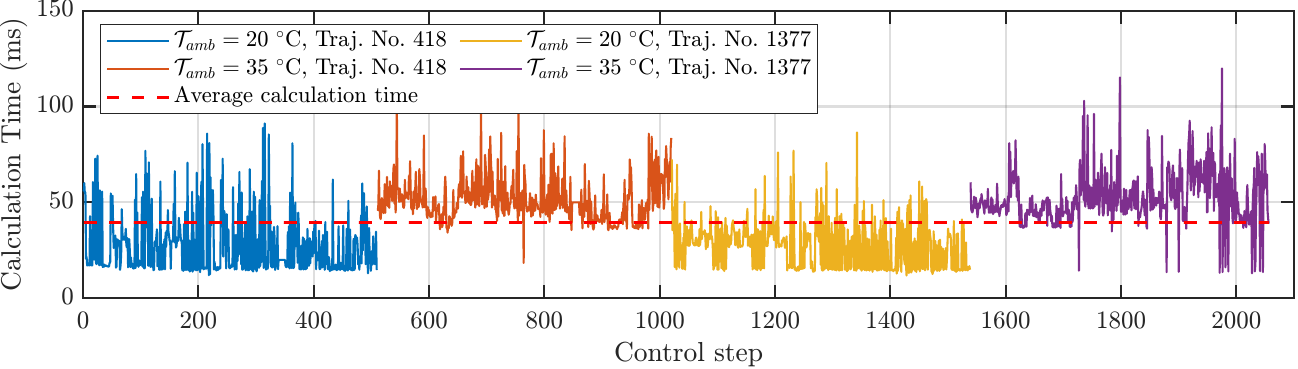}
  \caption{Calculation time of the proposed TA-ITEM strategy under different test scenarios.} \label{fig:Calculation_time}
\end{figure*}

Moreover, the computational efficiency of the proposed TA-ITEM strategy is evaluated under different vehicle trajectories and ambient temperatures. As shown in Fig.~\ref{fig:Calculation_time}, the total average computation time is 39.5 ms, demonstrating that the proposed strategy is suitable for real-time application \cite{DENG2021100094}. 
These results collectively demonstrate that TA-ITEM exhibits significant advantages in achieving coordinated energy-thermal optimization. Compared to conventional rule-based, MPC-SP, and two-layer strategies, it enhances energy efficiency, thermal regulation performance, and cabin comfort.

\section{Conclusion}\label{se:Conclusion}
This study proposes a traffic-aware hierarchical ITEM control approach for connected HEVs. By incorporating a Transformer-based speed prediction model with driving condition recognition, the framework enables anticipatory control that improves both energy efficiency and thermal comfort. The main findings are summarized as follows:
\begin{itemize}
  \item A control-oriented thermal model is developed and validated against high-fidelity simulation data, demonstrating high accuracy in capturing both transient and steady-state thermal behaviors.
  \item The proposed TF-DCR predictor improves speed prediction accuracy by up to 53.17\% compared to the ELM baseline.
  \item Comparative evaluations under both standard and real-world driving cycles show that TA-ITEM consistently outperforms rule-based and MPC-SP approaches, achieving average fuel savings of 56.36\% and 5.84\%, respectively, along with better battery thermal regulation and enhanced cabin thermal comfort.
\end{itemize}

These results confirm the strong generalization capability of the TA-ITEM framework and underscore the value of incorporating traffic foresight into multi-domain vehicle energy management. Future work will explore the extension of this approach to multi-vehicle coordination scenarios and its validation through real-time hardware-in-the-loop (HIL) experiments.


%





\ifCLASSOPTIONcaptionsoff
  \newpage
\fi





\bibliographystyle{IEEEtran}
\bibliography{IEEEabrv,Bibliography}

@IEEEtranBSTCTL{IEEEexample:BSTcontrol,
    CTLuse_article_number = "yes",
    CTLuse_paper = "yes",
    CTLuse_forced_etal = "no",
    CTLmax_names_forced_etal = "50",
    CTLnames_show_etal = "50",
    CTLuse_alt_spacing = "yes",
    CTLalt_stretch_factor = "4",
    CTLdash_repeated_names = "yes",
    CTLname_format_string = "{f.~}{vv~}{ll}{, jj}",
    CTLname_latex_cmd = "",
    CTLname_url_prefix = "[Online]. Available:"
 }

@article{han2023health,
  title={Health-conscious predictive energy management strategy with hybrid speed predictor for plug-in hybrid electric vehicles: Investigating the impact of battery electro-thermal-aging models},
  author={Han, Jie and Liu, Wenxue and Zheng, Yusheng and Khalatbarisoltani, Arash and Yang, Yalian and Hu, Xiaosong},
  journal={Applied Energy},
  volume={352},
  pages={121986},
  year={2023},
  publisher={Elsevier}
}

@article{herrera2010evaluation,
  title={Evaluation of traffic data obtained via GPS-enabled mobile phones: The Mobile Century field experiment},
  author={Herrera, Juan C and Work, Daniel B and Herring, Ryan and Ban, Xuegang Jeff and Jacobson, Quinn and Bayen, Alexandre M},
  journal={Transportation Research Part C: Emerging Technologies},
  volume={18},
  number={4},
  pages={568--583},
  year={2010},
  publisher={Elsevier}
}

@Article{Andersson2019,
  author = {Joel A E Andersson and Joris Gillis and Greg Horn
            and James B Rawlings and Moritz Diehl},
  title = {{CasADi} -- {A} software framework for nonlinear optimization
           and optimal control},
  journal = {Mathematical Programming Computation},
  volume = {11},
  number = {1},
  pages = {1--36},
  year = {2019},
  publisher = {Springer},
  doi = {10.1007/s12532-018-0139-4}
}

@article{chen2001freeway,
  title={Freeway performance measurement system: mining loop detector data},
  author={Chen, Chao and Petty, Karl and Skabardonis, Alexander and Varaiya, Pravin and Jia, Zhanfeng},
  journal={Transportation research record},
  volume={1748},
  number={1},
  pages={96--102},
  year={2001},
  publisher={SAGE Publications Sage CA: Los Angeles, CA}
}

@article{oh2020vehicle,
  title={Vehicle energy dataset (VED), a large-scale dataset for vehicle energy consumption research},
  author={Oh, Geunseob and Leblanc, David J and Peng, Huei},
  journal={IEEE Transactions on Intelligent Transportation Systems},
  volume={23},
  number={4},
  pages={3302--3312},
  year={2020},
  publisher={IEEE}
}

@misc{Autonomie2024,
  title        = {{Autonomie}: Vehicle System Simulation Tool},
  author       = {{Argonne National Laboratory, Vehicle \& Mobility Systems Department}},
  year         = 2024,
  url          = {https://vms.taps.anl.gov/tools/autonomie/},
  note         = {Accessed: 2025-06-12},
  organization = {Argonne National Laboratory},
}

@ARTICLE{6930758,
  author={Sun, Chao and Moura, Scott Jason and Hu, Xiaosong and Hedrick, J. Karl and Sun, Fengchun},
  journal={IEEE Transactions on Control Systems Technology}, 
  title={Dynamic Traffic Feedback Data Enabled Energy Management in Plug-in Hybrid Electric Vehicles}, 
  year={2015},
  volume={23},
  number={3},
  pages={1075-1086},
  doi={10.1109/TCST.2014.2361294}}

@online{caltrans_pems,
  author       = {California Department of Transportation},
  title        = {Performance Measurement System (PeMS)},
  year         = {2025},
  note         = {Accessed: Jun. 2025},
  url          = {https://pems.dot.ca.gov/}
}

@article{tang2020naturalistic,
  title={Naturalistic data-driven predictive energy management for plug-in hybrid electric vehicles},
  author={Tang, Xiaolin and Jia, Tong and Hu, Xiaosong and Huang, Yanjun and Deng, Zhongwei and Pu, Huayan},
  journal={IEEE Transactions on Transportation Electrification},
  volume={7},
  number={2},
  pages={497--508},
  year={2020},
  publisher={IEEE}
}

@article{han2022predictive,
  title={Predictive energy management for plug-in hybrid electric vehicles considering electric motor thermal dynamics},
  author={Han, Jie and Shu, Hong and Tang, Xiaolin and Lin, Xianke and Liu, Changpeng and Hu, Xiaosong},
  journal={Energy Conversion and Management},
  volume={251},
  pages={115022},
  year={2022},
  publisher={Elsevier}
}

@article{shams2012integrated,
  title={Integrated thermal and energy management of plug-in hybrid electric vehicles},
  author={Shams-Zahraei, Mojtaba and Kouzani, Abbas Z and Kutter, Steffen and B{\"a}ker, Bernard},
  journal={Journal of power sources},
  volume={216},
  pages={237--248},
  year={2012},
  publisher={Elsevier}
}

@article{hu2021multihorizon,
  title={Multihorizon model predictive control: An application to integrated power and thermal management of connected hybrid electric vehicles},
  author={Hu, Qiuhao and Amini, Mohammad Reza and Kolmanovsky, Ilya and Sun, Jing and Wiese, Ashley and Seeds, Julia Buckland},
  journal={IEEE Transactions on Control Systems Technology},
  volume={30},
  number={3},
  pages={1052--1064},
  year={2021},
  publisher={IEEE}
}

@article{wu2024integrated,
  title={Integrated battery thermal and energy management for electric vehicles with hybrid energy storage system: A hierarchical approach},
  author={Wu, Yue and Huang, Zhiwu and Li, Dongjun and Li, Heng and Peng, Jun and Guerrero, Josep M and Song, Ziyou},
  journal={Energy Conversion and Management},
  volume={317},
  pages={118853},
  year={2024},
  publisher={Elsevier}
}

@article{khalatbarisoltani2025privacy,
  title={Privacy-preserving integrated thermal and energy management of multi connected hybrid electric vehicles with federated reinforcement learning},
  author={Khalatbarisoltani, Arash and Han, Jie and Saeed, Muhammad and Liu, Cong-zhi and Hu, Xiaosong},
  journal={Applied Energy},
  volume={385},
  pages={125386},
  year={2025},
  publisher={Elsevier}
}

@article{zhang2023integrated,
  title={Integrated thermal and energy management of connected hybrid electric vehicles using deep reinforcement learning},
  author={Zhang, Hao and Chen, Boli and Lei, Nuo and Li, Bingbing and Li, Rulong and Wang, Zhi},
  journal={IEEE Transactions on Transportation Electrification},
  volume={10},
  number={2},
  pages={4594--4603},
  year={2023},
  publisher={IEEE}
}

@article{zhao2021two,
  title={A two-layer real-time optimization control strategy for integrated battery thermal management and HVAC system in connected and automated HEVs},
  author={Zhao, Shuofeng and Amini, Mohammad Reza and Sun, Jing and Mi, Chunting Chris},
  journal={IEEE Transactions on Vehicular Technology},
  volume={70},
  number={7},
  pages={6567--6576},
  year={2021},
  publisher={IEEE}
}

@article{amini2019cabin,
  title={Cabin and battery thermal management of connected and automated HEVs for improved energy efficiency using hierarchical model predictive control},
  author={Amini, Mohammad Reza and Wang, Hao and Gong, Xun and Liao-McPherson, Dominic and Kolmanovsky, Ilya and Sun, Jing},
  journal={IEEE Transactions on Control Systems Technology},
  volume={28},
  number={5},
  pages={1711--1726},
  year={2019},
  publisher={IEEE}
}

@article{dong2024predictive,
  title={Predictive battery thermal and energy management for connected and automated electric vehicles},
  author={Dong, Haoxuan and Hu, Qiuhao and Li, Dongjun and Li, Zhaojian and Song, Ziyou},
  journal={IEEE Transactions on Intelligent Transportation Systems},
  year={2024},
  publisher={IEEE}
}

@article{han2023energy,
  title={Energy management in plug-in hybrid electric vehicles: preheating the battery packs in low-temperature driving scenarios},
  author={Han, Jie and Khalatbarisoltani, Arash and Yang, Yalian and Hu, Xiaosong},
  journal={IEEE Transactions on Intelligent Transportation Systems},
  volume={25},
  number={2},
  pages={1978--1991},
  year={2023},
  publisher={IEEE}
}

@article{liang2023efficient,
  title={Efficient mode transition control for DM-PHEV with mechanical hysteresis based on piecewise affine H$\infty$ strategy},
  author={Liang, Cong and Xu, Xing and Auger, Daniel J and Wang, Feng and Wang, Shaohua},
  journal={IEEE Transactions on Transportation Electrification},
  volume={9},
  number={3},
  pages={4366--4379},
  year={2023},
  publisher={IEEE}
}

@article{wei2019integrated,
  title={Integrated energy and thermal management for electrified powertrains},
  author={Wei, Caiyang and Hofman, Theo and Ilhan Caarls, Esin and van Iperen, Rokus},
  journal={Energies},
  volume={12},
  number={11},
  pages={2058},
  year={2019},
  publisher={MDPI}
}

@inproceedings{gong2019integrated,
  title={Integrated optimization of power split, engine thermal management, and cabin heating for hybrid electric vehicles},
  author={Gong, Xun and Wang, Hao and Amini, Mohammad Reza and Kolmanovsky, Ilya and Sun, Jing},
  booktitle={2019 IEEE Conference on Control Technology and Applications (CCTA)},
  pages={567--572},
  year={2019},
  organization={IEEE}
}

@article{he2023review,
  title={Review of thermal management system for battery electric vehicle},
  author={He, Liange and Jing, Haodong and Zhang, Yan and Li, Pengpai and Gu, Zihan},
  journal={Journal of Energy Storage},
  volume={59},
  pages={106443},
  year={2023},
  publisher={Elsevier}
}

@article{lu2021combined,
  title={A combined method for short-term traffic flow prediction based on recurrent neural network},
  author={Lu, Saiqun and Zhang, Qiyan and Chen, Guangsen and Seng, Dewen},
  journal={Alexandria Engineering Journal},
  volume={60},
  number={1},
  pages={87--94},
  year={2021},
  publisher={Elsevier}
}

@article{pulvirenti2023energy,
  title={Energy management system optimization based on an LSTM deep learning model using vehicle speed prediction},
  author={Pulvirenti, Luca and Rolando, Luciano and Millo, Federico},
  journal={Transportation Engineering},
  volume={11},
  pages={100160},
  year={2023},
  publisher={Elsevier}
}

@article{zhu2025vspnet,
  title={VSPNet: a vehicle speed prediction model incorporating transformer and BiLSTM},
  author={Zhu, Qinglin and Chen, Dehui and Wang, Zhangu and Lv, Baibing and Zhao, Ziliang and Zhao, Jun},
  journal={Measurement Science and Technology},
  volume={36},
  number={2},
  pages={026118},
  year={2025},
  publisher={IOP Publishing}
}

@article{quan2021real,
  title={Real-time energy management for fuel cell electric vehicle using speed prediction-based model predictive control considering performance degradation},
  author={Quan, Shengwei and Wang, Ya-Xiong and Xiao, Xuelian and He, Hongwen and Sun, Fengchun},
  journal={Applied Energy},
  volume={304},
  pages={117845},
  year={2021},
  publisher={Elsevier}
}

@article{yang2023modeling,
  title={Modeling and simulation of vehicle integrated thermal management system for a fuel cell hybrid vehicle},
  author={Yang, Qiao and Zeng, Tao and Zhang, Caizhi and Zhou, Wenjian and Xu, Lei and Zhou, Jiaming and Jiang, Ping and Jiang, Shangfeng},
  journal={Energy Conversion and Management},
  volume={278},
  pages={116745},
  year={2023},
  publisher={Elsevier}
}

@inproceedings{lokur2023distributed,
  title={Distributed Model Predictive Controller For Thermal Energy Management System of Battery Electric Vehicles},
  author={Lokur, Prashant and Murgovski, Nikolce and Nicklasson, Kristian},
  booktitle={2023 62nd IEEE Conference on Decision and Control (CDC)},
  pages={8363--8368},
  year={2023},
  organization={IEEE}
}

@article{amini2020hierarchical,
  title={Hierarchical MPC for robust eco-cooling of connected and automated vehicles and its application to electric vehicle battery thermal management},
  author={Amini, Mohammad Reza and Kolmanovsky, Ilya and Sun, Jing},
  journal={IEEE Transactions on Control Systems Technology},
  volume={29},
  number={1},
  pages={316--328},
  year={2020},
  publisher={IEEE}
}

@article{hu2022multirange,
  title={A multirange vehicle speed prediction with application to model predictive control-based integrated power and thermal management of connected hybrid electric vehicles},
  author={Hu, Qiuhao and Amini, Mohammad Reza and Wiese, Ashley and Seeds, Julia Buckland and Kolmanovsky, Ilya and Sun, Jing},
  journal={Journal of Dynamic Systems, Measurement, and Control},
  volume={144},
  number={1},
  pages={011105},
  year={2022},
  publisher={American Society of Mechanical Engineers}
}

@article{zhou2022comprehensive,
  title={A comprehensive study of speed prediction in transportation system: From vehicle to traffic},
  author={Zhou, Zewei and Yang, Ziru and Zhang, Yuanjian and Huang, Yanjun and Chen, Hong and Yu, Zhuoping},
  journal={Iscience},
  volume={25},
  number={3},
  year={2022},
  publisher={Elsevier}
}

@article{lin2021velocity,
  title={Velocity prediction using Markov Chain combined with driving pattern recognition and applied to Dual-Motor Electric Vehicle energy consumption evaluation},
  author={Lin, Xinyou and Zhang, Guangji and Wei, Shenshen},
  journal={Applied Soft Computing},
  volume={101},
  pages={106998},
  year={2021},
  publisher={Elsevier}
}

@article{li2020driver,
  title={Driver-identified supervisory control system of hybrid electric vehicles based on spectrum-guided fuzzy feature extraction},
  author={Li, Ji and Zhou, Quan and He, Yinglong and Williams, Huw and Xu, Hongming},
  journal={IEEE Transactions on Fuzzy Systems},
  volume={28},
  number={11},
  pages={2691--2701},
  year={2020},
  publisher={IEEE}
}

@incollection{li2025driver,
  title={Driver Behavior Prediction and Driver-Oriented Control of Electric Vehicles},
  author={Li, Ji and Zhou, Quan and Wu, Yanhong and Xu, Hongming},
  booktitle={Big Data and Electric Mobility},
  pages={171--189},
  year={2025},
  publisher={CRC Press}
}

@article{han2022joint,
  title={Joint optimization of configuration, component sizing, and energy management for input-split hybrid powertrains},
  author={Han, Jie and Hu, Xiaosong and Yang, Yalian and Grzesiak, Lech M and Kum, Dongsuk},
  journal={IEEE Transactions on Vehicular Technology},
  volume={72},
  number={2},
  pages={1649--1661},
  year={2022},
  publisher={IEEE}
}

@book{onori2016hybrid,
  title={Hybrid electric vehicles: Energy management strategies},
  author={Onori, Simona and Serrao, Lorenzo and Rizzoni, Giorgio},
  volume={13},
  year={2016},
  publisher={Springer}
}

@article{bonab2020fuel,
  title={Fuel-optimal energy management strategy for a power-split powertrain via convex optimization},
  author={Bonab, Saeed Amirfarhangi and Emadi, Ali},
  journal={IEEE Access},
  volume={8},
  pages={30854--30862},
  year={2020},
  publisher={IEEE}
}

@article{DENG2021100094,
title = {An adaptive PMP-based model predictive energy management strategy for fuel cell hybrid railway vehicles},
journal = {eTransportation},
volume = {7},
pages = {100094},
year = {2021},
issn = {2590-1168},
author = {Kai Deng and Hujun Peng and Steffen Dirkes and Jonas Gottschalk and Cem Ünlübayir and Andreas Thul and Lars Löwenstein and Stefan Pischinger and Kay Hameyer},
}

\vfill


\end{document}